\documentclass[12pt,dvips]{article}
\usepackage{here}
\usepackage{amsbsy,amssymb}
\newcommand{\be}[1]{\begin{equation}\label{#1}}
\newcommand{\ee}{\end{equation}}
\newcommand{\beq}[1]{\begin{eqnarray}\label{#1}}
\newcommand{\eeq}{\end{eqnarray}}
\newcommand{\ba}{\begin{array}}
\newcommand{\ea}{\end{array}}
\newcommand{\of}[1]{\left(#1\right)}
\newcommand{\off}[1]{\left[#1\right]}
\newcommand{\offf}[1]{\left\{#1\right\}}

\newcommand{\gb}[2]{\off{\!\off{#1,#2}\!}}
\newcommand{\bgb}[2]{\off{\!\!\off{#1,#2}\!\!}}
\newcommand{\bs}[1]{\boldsymbol{#1}}
\newcommand{\ms}[1]{\mbox{\small#1}}
\newcommand{\mf}[1]{\mbox{\footnotesize#1}}

\newcommand{\Par}{\par\vspace{0.2cm}\par}
\newcommand{\R}{{\cal R}}

\newcommand{\F}{{\cal F}}
\newcommand{\B}{{\cal B}}
\newcommand{\D}{{\cal D}}

\newcommand{\gl}{\mathfrak{g}\mathfrak{l}}

\renewcommand{\baselinestretch}{1.6}\tiny\normalsize
\begin{document}
\title{\renewcommand{\baselinestretch}{2.2}\tiny\normalsize
{\Large\sf Non-Split Geometry on Products of Vector
Bundles}
\author{{\sc O. Megged}\thanks{e-mail address: 
megged@post.tau.ac.il}\\
{\small\rm School of Physics and Astronomy}\\
{\small\rm Tel-Aviv University, Tel-Aviv 69978, Israel.}}}
\date{}
\maketitle
\begin{abstract}
We propose a model in which a spliced vector bundle (with an arbitrary 
number of gauge structures in the splice) possesses a geometry which do
not split. The model employs connection 1-forms with values in a
space-product of Lie algebras, and therefore interlaces the various
gauge structures in a non-trivial manner. Special attention is given
to the structure of the geometric ghost sector and the super-algebra
it possesses: The ghosts emerge as $x$-dependent deformations at the
gauge sector, and the associated BRST super algebra is realized as
constraints that follow from the invariance of the curvature.
\end{abstract}
%-------------
\section{Introduction}\label{S1}
%-------------
A product of vector bundles, within the classical framework of gauge 
theories, is often contemplated as the bundle of product-space fibers, 
where each factor-fiber in the splice is a representation space for a
certain gauge group. This results in a geometrical splitting by the
following means: When the  
geometrical aspects of a single group structure are considered, those 
components of a geometric object that correspond to other coexisting 
group structures, all remain non-active. This fact is after all a
consequence of the Leibniz rule. For example: The absolute
differential of a tensor product of two fiber bases splits into a sum
of tensor products of single-basis differentials, which are in turn
used to define the corresponding factor structure connections,  
\beq{1}
d\of{\bs{e}_{1}\otimes\bs{e}_{2}}&=&
\of{d\bs{e}_{1}}\otimes\bs{e}_{2}+
\bs{e}_{1}\otimes d\of{\bs{e}_{2}}
\nonumber\\&=:&
-\omega_{1}\of{\bs{e}_{1}}\otimes\bs{e}_{2}-
\bs{e}_{1}\otimes\omega_{2}\of{\bs{e}_{2}}.
\eeq\Par
This splitting, however, is not compatible with the concepts
of fusion and unification. The following 
question therefore arises:  Is it possible to form a better glue of
symmetry structures, a one that results in a single non-split geometry
of the composite bundle? In this article we provide 
an affirmative answer to this question. We shall replace definition
(\ref{1}) with a somewhat less intuitive definition,
\beq{2}
d\of{\bs{e}_{1}\otimes\bs{e}_{2}}&=&
\of{d\bs{e}_{1}}\otimes\bs{e}_{2}+
\bs{e}_{1}\otimes d\of{\bs{e}_{2}}
\nonumber\\&=:&
-\omega_{1}\of{\bs{e}_{1}\otimes\bs{e}_{2}}
-\omega_{2}\of{\bs{e}_{1}\otimes\bs{e}_{2}},
\eeq
and discuss the conditions under which it is really meaningful. This
will lead to a geometry which do not split even though the
bundle itself inherently splits.\Par
Our model is based on a collection of connection 1-forms, each
taking values in a space product of Lie algebras, instead of in a
single Lie algebra. These are later integrated to form a single
curvature of the multi-structure splice, and the latter obeys the
Bianchi identity with respect to an appropriately-constructed
covariant exterior derivative.\Par 
We shall also derive the associated ghost structure and BRST
symmetry by pure geometrical means: The connections undergo a (local)
deformation, and the basespace is extended by multiplying it (locally
again) with the spaces spanned by the symmetry groups. The exterior
derivatives in group-spaces are then identified with
the BRST coboundary operators, and the deformation elements at the
the gauge sector are identified with the ghosts. The BRST algebra
then emerges as structural constraints that follow directly from the
demand that the curvature of the whole splice will remains intact.\Par
We refer the reader to the following physically-inclined
accounts as a background material for this article:
A concise presentation
of the concept of fiber bundles is given in \cite{B} p. 95-117; the
concept of a product bundle, where distinct symmetry structures share
a common basespace, is discussed in \cite{NS} p. 194-196. A
detailed presentation and analysis of ghosts and BRST symmetries from
the field theory point of view is found in \cite{NO} p. 141-181. The
same subject, presented from the geometrical perspective (more
relevant to our proposes) is given in \cite{B}, chapters 8 \& 9. 
\Par\Par\noindent
{\em  Notations and Conventions\/}:\Par\noindent
We shall consider a product bundle that hosts an arbitrary (but finite)
number of coexisting symmetry structures, say $m$.
The underlying manifold $M$ is smooth and
oriented. $\B_{x}$ is the basespace at $x$, $\F_{x}$ is the local 
fiber. It consists of a product space of $m$ vector spaces
$\offf{V_{\alpha}}$, each of which is acted upon by a specific
gauge group $G_{\alpha}\of{x}$. Here and in the following,
$\alpha,\gamma,\ldots=1,\ldots,m$ label the members of the hosted vector 
spaces, the associated symmetry groups, and their generating  Lie
algebras; $n=\mbox{dim}\,M$, $n_{\alpha}=\mbox{dim}\,G_{\alpha}$,  
$N_{\alpha}=\mbox{dim}\,V_{\alpha}$.\Par
Our conventions with respect to indices goes as follows:
$a_{\alpha},b_{\alpha},\ldots=1\cdots n_{\alpha}$ are 
$G_{\alpha}$-indices (associated with the symmetry group $G_{\alpha}$).
$A_{\alpha},B_{\alpha},\ldots=1\cdots  
N_{\alpha}$ are $V_{\alpha}$-fiberspace indices (associated with the
representation space $V_{\alpha}$ of $G_{\alpha}$). The
basespace employs Greek indices, $\mu,\nu,\ldots=1\cdots n$ which are,
without lose of generality, taken to be holonomic. Concerning with
brackets notation, for any $p$-form $\psi$, and $q$-form $\phi$,
\beq{S8}
\off{\psi,\phi}_{\mp}&:=&\psi\wedge\phi\mp\phi\wedge\psi,\\
\gb{\psi}{\phi}&:=&\psi\wedge\phi-\of{-1}^{pq}\phi\wedge\psi.
\eeq
%-------------
\section{The foliar bundle and its associated curvature}\label{S2}
%-------------
Let $F$ refer to a product bundle which consists of $m$ distinct
independent coexisting gauge structures. The elements
$\offf{L^{\alpha}}$ of the algebra $\mbox{Lie}\,G_{\alpha}$ that
generates $G_{\alpha}$ (for any $\alpha=1\cdots{m}$), are assumed
to carry a faithful representation $\rho_{\alpha}$ in $V_{\alpha}$,
and to extend to the full enveloping algebra, so their anti-commutator
is well defined. In what follows we 
shall restrict ourselves to deal only with $G_{\alpha}$-structures
that possess representations in which 
\be{S10}
\off{\rho_{\alpha}\!\of{L^{\alpha}},
\rho_{\alpha}\!\of{L^{\alpha}}}_{+}
\in\;\mbox{span}\offf{\rho_{\alpha}\of{L^{\alpha}}};
\ee
namely, {\em the realizations of the algebras close with respect to anti-commutation}. 
An algebra whose elements in a certain representation close with
respect to anti-commutation is said to be {\em sealed} in that
representation. The requirement that the algebra be sealed in a
representation is obligatory to our present purposes. A simple
example of such an algebra is the one which generates invertible
linear transformations in a vector space,
$\gl\of{n,\mathbb{R}}$. Concerning with unitary gauge
structures, the closure relation (\ref{S10}), can be realized only
if the algebra is extended by central elements.\Par 
Let us now introduce a set of $m$ $G_{\alpha}$-induced connection 1-forms
$\offf{\omega_{\alpha}}$ with values in the {\em symmetric\/} product-space
$\bigotimes_{\alpha=1}^{m}\of{\mbox{Lie}\,G_{\alpha}}$,
\beq{S20}
\omega_{\alpha}=:
\rho_{F}\of{\widetilde{\omega}_{\alpha}}
=:\widetilde{\omega}_{\mu}^{a_{1}\cdots a_{m}}\of{G_{\alpha}}dx^{\mu}
\rho_{a_{1}\cdots a_{m}}\,,
\eeq
where $\offf{\widetilde{\omega}^{a_{1}\cdots a_{m}}
\of{G_{\alpha}}}$ are those connection coefficients which are
associated with the gauge group $G_{\alpha}$, the short-hand
writing $\rho_{a_{1}\cdots a_{m}}$ stands for the (symmetric)
product of matrices,  
\be{S30}
\of{\rho_{a_{1}\cdots a_{m}}}_{A_{1}\cdots A_{m}}^{B_{1}\cdots B_{m}}:=
\bigotimes_{\alpha=1}^{m}\,\of{\rho_{\alpha}
\of{L^{\alpha}_{a_{\alpha}}}}{\!}^{B_{\alpha}}_{{A_{\alpha}}},
\ee
and we defined $\rho_{F}\of{\varpi}:=
\varpi^{a_{1}\cdots{a}_{m}}\rho_{a_{1}\cdots{a}_{m}}$ for any
$\varpi\in\bigotimes_{\alpha=1}^{m}\of{\mbox{Lie}\,G_{\alpha}}$.
Notice that, in general, 
${\widetilde{\omega}^{a_{1}\cdots a_{m}}
\of{G_{\alpha}}}\neq{\widetilde{\omega}^{a_{1}\cdots a_{m}}
\of{G_{\gamma}}}$ for $\alpha\neq\gamma$, hence
$\omega_{\alpha}\neq\omega_{\gamma}$.\Par 
Under a gauge transformation (see also the discussion concerning with
eq. (\ref{S100})), each element in the collection
$\offf{\omega_{\alpha}}$ should transform as: 
\be{S40}
\forall\; g_{\gamma}\in G_{\gamma}:\left\{
\ba{llll}
\omega_{\alpha} & \mapsto &
g_{\gamma}\of{\omega_{\alpha}+d}g_{\gamma}^{-1} &
\;\;\gamma=\alpha \\
\omega_{\alpha} & \mapsto &
g_{\gamma}\omega_{\alpha}g_{\gamma}^{-1} &
\;\;\gamma\neq\alpha 
\ea\right.\ee
where the actions of the $g$'s are given by means of matrix
multiplication. Each $\omega_{\alpha}$, therefore,
transforms as a connection with respect to its inducing gauge group 
$G_{\alpha}$, but it behaves as a tensor with respect to the
rest of the groups in the collection. \Par
The set of connection 1-forms introduced above is seen to give rise to a
unique curvature 2-form which characterizes the whole splice:
\be{S50}
\R_{F}\of{\omega_{1},\cdots,\omega_{m}}=\sum_{\alpha,\gamma=1}^{m}
\of{d\omega_{\alpha}+\omega_{\alpha}\wedge\omega_{\gamma}} .
\ee
To see that this is indeed a ``proper'' curvature, we follow two
steps:\Par 
First we verify that $\R_{F}$ as well takes values
$\in\bigotimes_{\alpha}\of{\mbox{Lie}\,G_{\alpha}}$. 
This however follows directly from our previous demand, eq. (\ref{S10}),
that $\rho_{\alpha}\of{\mbox{Lie}\,G_{\alpha}}$ (for any
$\alpha=1\cdots{m}$) closes with respect to anti-commutation. In this case,
\beq{S60}
\rho_{\alpha}\of{L^{\alpha}}\rho_{\alpha}\of{L^{\alpha}}
&=&
\frac{1}{2}\off{\rho_{\alpha}\of{L^{\alpha}},\rho_{\alpha}\of{L^{\alpha}}}_{+}
+\frac{1}{2}\off{\rho_{\alpha}\of{L^{\alpha}},\rho_{\alpha}\of{L^{\alpha}}}_{-}
\;\subset\;\mbox{span}\offf{\rho_{\alpha}\of{\mbox{Lie}\,G_{\alpha}}}.
\nonumber\\
\eeq
Since each term in
$\sum_{\alpha,\gamma}\omega_{\alpha}\wedge\omega_{\gamma}$ contains
products of pairs of generators (the elements in each pair
belong to the same Lie algebra), assignment (\ref{S60}) guarantees
that the resulting algebraic expansion will always lay in
$\bigotimes_{\alpha}\of{\mbox{Lie}\,G_{\alpha}}$. Hence, 
\beq{S70}
\R_{F}&:=&
\R_{\mu\nu}\of{\rho_{F}\of{\widetilde{\omega}}}dx^{\mu}\wedge{d}x^{\nu}
=\sum_{\alpha,\gamma=1}^{m}\off{d\rho_{F}\of{\widetilde{\omega}_{\alpha}}+
\rho_{F}\of{\widetilde{\omega}_{\alpha}}\wedge
\rho_{F}\of{\widetilde{\omega}_{\gamma}}}\nonumber\\
&\!\!=\!\!&\sum_{\alpha,\gamma=1}^{m}
d\rho_{F}\of{\widetilde{\omega}_{\alpha}}+\rho_{F}\of{
\widetilde{\omega}_{\alpha}\wedge\widetilde{\omega}_{\gamma}}\,=\,
\rho_{F}\of{\R_{\mu\nu}\of{\widetilde{\omega}}}dx^{\mu}\wedge{dx}^{\nu},
\eeq
where $\rho_{F}\of{\R}=
\R^{a_{1}\cdots{a}_{m}}\rho_{a_{1}\cdots{a}_{m}}$.
We shall adopt the shortage notation: $a_{1}\cdots{a}_{m}=\offf{a}$ etc.
Then, $\rho_{F}\of{\R}=\R^{\offf{a}}\rho_{\offf{a}}$, and 
\be{S80}
\R^{\offf{a}}\,=\,\sum_{\alpha=1}^{m}
d\widetilde{\omega}_{\alpha}^{\offf{a}}+
\sum_{\alpha,\gamma=1}^{m}
f^{\offf{a}\offf{b}\offf{c}}\,
\widetilde{\omega}_{\alpha}^{\offf{b}}\wedge
\widetilde{\omega}_{\gamma}^{\offf{c}},
\ee
where $f^{\offf{a}\offf{b}\offf{c}}$ is defined through 
$\off{\rho_{\offf{a}},\rho_{\offf{b}}}_{-}=
f_{\offf{a}\offf{b}\offf{c}}\rho_{\offf{c}}$.\footnote{
Formula (\ref{S80}) is coming from:
\beq{S80a}
\sum_{\alpha,\gamma}\omega_{\alpha}\wedge\omega_{\gamma}&=&
\frac{1}{2}\sum_{\alpha,\gamma}
\sum_{\offf{a}}^{\offf{n}}
\sum_{\offf{b}}^{\offf{n}}
\off{
\of{\widetilde\omega_{\alpha}}_{\mu}^{\offf{a}}\rho_{\offf{a}}
\of{\widetilde\omega_{\gamma}}_{\nu}^{\offf{b}}\rho_{\offf{b}}
-
\of{\widetilde\omega_{\alpha}}_{\nu}^{\offf{a}}\rho_{\offf{a}}
\of{\widetilde\omega_{\gamma}}_{\mu}^{\offf{b}}\rho_{\offf{b}}
}
dx^{\mu}\wedge{dx}^{\nu}
\nonumber\\&=&
\frac{1}{2}
\sum_{\alpha,\gamma}
\sum_{\offf{a}}^{\offf{n}}
\sum_{\offf{b}}^{\offf{n}}
\of{\widetilde\omega_{\alpha}}_{\mu}^{\offf{a}}
\of{\widetilde\omega_{\gamma}}_{\nu}^{\offf{b}}
\off{\rho_{\offf{a}},\rho_{\offf{b}}}_{-}dx^{\mu}\wedge{dx}^{\nu}
\nonumber\\&=&
\sum_{\alpha,\gamma}
\sum_{\offf{a}}^{\offf{n}}
\sum_{\offf{b}}^{\offf{n}}
\sum_{\offf{c}}^{\offf{n}}
\widetilde\omega_{\alpha}^{\offf{a}}\wedge
\widetilde\omega_{\gamma}^{\offf{b}}
f_{\offf{a}\offf{b}\offf{c}}\rho_{\offf{c}}.
\eeq}
\Par
Second, we should verify that $\R_{F}$ transforms linearly (namely,
as a tensor) with respect to each gauge group in the collection.
Indeed, considering a particular label, say $\alpha$, $\R_{F}$ can 
be decomposed as,
\beq{S90a}
&&d\omega_{\alpha}+\omega_{\alpha}\wedge\omega_{\alpha}\nonumber\\
&+&\sum_{\gamma\neq\alpha}\of{d\omega_{\gamma}+
\omega_{\alpha}\wedge\omega_{\gamma}+
\omega_{\gamma}\wedge\omega_{\alpha}}\nonumber\\
&+&\sum_{\gamma,\epsilon\neq\alpha}\omega_{\gamma}\wedge\omega_{\epsilon}
\eeq
which we may also write as
\be{S90b}
\R_{F}\,=\,\R_{\alpha}\of{\omega_{\alpha}}
+\sum_{\gamma\neq\alpha}\D_{\omega_{\alpha}}\omega_{\gamma}
+\sum_{\gamma,\epsilon\neq\alpha}\omega_{\gamma}\wedge\omega_{\epsilon},
\ee
where $\D_{\omega_{\alpha}}\omega_{\gamma}$ is the covariant exterior
derivative of $\omega_{\gamma}$ with respect to the connection
$\omega_{\alpha}$. 
Under the action of $G_{\alpha}$, each summand in (\ref{S90b}) transforms
linearly, and in a manner which is independent of all the other summands,
\Par\noindent
\be{S92}
\forall\;g_{\alpha}\in{G}_{\alpha}\;\;\left\{\;
\ba{rclcl}
R_{\alpha}\of{\omega_{\alpha}}&\mapsto&
g_{\alpha}R_{\alpha}\of{\omega_{\alpha}}g_{\alpha}^{-1},&&\\
{D_{\omega_{\alpha}}\omega_{\gamma}}&\mapsto&
g_{\alpha}\of{D_{\omega_{\alpha}}{\omega_{\gamma}}}g_{\alpha}^{-1}
&&\of{\gamma\neq\alpha},\\
\omega_{\gamma}\wedge\omega_{\epsilon}&\mapsto&
g_{\alpha}\of{\omega_{\gamma}\wedge\omega_{\epsilon}}g_{\alpha}^{-1}&&
\of{\gamma,\epsilon\neq\alpha}.
\ea\right.
\ee\Par\noindent
This, however, holds for any $\alpha=1,\ldots,m$; thus $\R_{F}$ is
linear with respect to all the $G$'s and our claim that $\R_{F}$ is a
proper curvature has been established.\Par 
In fact, ${\sum_{\alpha}\omega_{\alpha}}:=\omega_{F}$ can
be regarded as a single connection, having the property of
simultaneously supporting many gauges:
\be{S94}
\ba{lrcl}
\forall\;\alpha\;\&\;\forall\; 
g_{\alpha}\in G_{\alpha}:\;\;&
\omega_{F}&\mapsto&{g}_{\alpha}\of{\omega_{F}+d}g_{\alpha}^{-1}.
\ea
\ee
Therefore, $\omega_{F}$ underlies a generic formation
of gauge, in which $m$ distinct coexisting structures are 
intertwined, and whose associated curvature acquires
a `single-structure' form, 
\be{S96}
\R_{F}=d\omega_{F}+\omega_{F}\wedge\omega_{F}.
\ee 
We therefore name our geometrical construction ``{\em foliar bundle\/}''; it is 
a product-bundle whose single-foil slices are interlaced in such a way 
that the overall geometry do not split.\Par
The set of connection 1-forms introduced in (\ref{S20}) can naturally 
be derived by considering the absolute differential of the multi-foil
basis, $\bs{e}_{A_{1}\cdots A_{m}}=
\bigotimes_{\alpha=1}^{m}\bs{e}^{\alpha}_{A_{\alpha}}$:
\beq{S100}
d\bs{e}_{A_{1}\cdots A_{m}}&\!=\!&
d\bs{e}_{A_{1}}^{1}\otimes\cdots\otimes\bs{e}_{A_{m}}^{m}\,+\;
\cdots\;+\;
\bs{e}_{A_{1}}^{1}\otimes\cdots\otimes{d}\bs{e}_{A_{m}}^{m}
\nonumber\\&:=&
\sum_{\alpha=1}^{m}
-\rho_{F}
\of{\widetilde{\omega}_{\alpha}}^{B_{1}\cdots B{m}}_{A_{1}\cdots A_{m}}
\bs{e}_{B_{1}\cdots B_{m}},
\eeq
which is conveniently abbreviated as
$d\bs{e}_{A_{1}\cdots A_{m}}=-\sum_{\alpha}\of{\omega_{\alpha}
\bs{e}}_{A_{1}\cdots A_{m}}\,$. Definition (\ref{S100}) remains valid
in any gauge (in any symmetry slice) provided that the
$\omega_{\alpha}$'s transforms as in eq. (\ref{S40}).\Par
Additional application of $d$ on the multi-foil basis gives
\beq{S120}
dd\bs{e}_{A_{1}\cdots A_{m}}&=&
\off{-\sum_{\alpha}
d\rho_{F}\of{\widetilde{\omega}_{\alpha}}
_{A_{1}\cdots{A}_{m}}^{B_{1}\cdots{B}_{m}}
-\sum_{\alpha}\sum_{\gamma}
\rho_{F}\of{\widetilde\omega_{\alpha}}
_{A_{1}\cdots{A}_{m}}^{C_{1}\cdots{C}_{m}}
\wedge\rho_{F}\of{\widetilde\omega_{\gamma}}
_{C_{1}\cdots{C}_{m}}^{B_{1}\cdots{B}_{m}}}
\bs{e}_{B_{1}\cdots{B}_{m}}\nonumber\\
&=&-\of{\R_{F}\bs{e}}_{A_{1}\cdots A_{m}}.
\eeq\Par
Consider the $\of{{\bs{\times}}_{\gamma}G_{\gamma}}$-tensor object
$\Psi_{T}$ (it transforms as a tensor with respect to any of
the $G$'s). By formula (\ref{S40}), its (foliar)
covariant exterior derivative, 
\beq{S160}
\D\Psi_{T}&:=&d\Psi_{T}+\sum_{\alpha}\of{\omega_{\alpha}\wedge\Psi_{T}+
\of{-1}^{\ms{deg}\of{\Psi_{T}}+1}\Psi_{T}\wedge\omega_{\alpha}}\\
&=&d\Psi_{T}+\omega_{F}\wedge\Psi_{T}+\of{-1}^{\ms{deg}\of{\Psi_{T}}+1}
\Psi_{T}\wedge\omega_{F}\;=\;
d\Psi_{T}+\gb{\omega_{F}}{\Psi_{T}}\label{S162}
\eeq
is a $\of{{\bs{\times}}_{\gamma}G_{\gamma}}$-tensor as well; namely
$\D\Psi_{T}$ transforms as a tensor with respect to any of the 
$G$'s. In terms of this covariant exterior derivative, the foliar
curvature 
can be re-constructed via $\D\D\Psi_{T}=\off{\R_{F},\Psi_{T}}_{-}$. 
Moreover, by the (graded) Jacobi identity,
\beq{S170}
0\;=\;\gb{\D}{\gb{\D}{\D}}\Psi_{T}&=&2\D\off{\R_{F},\Psi_{T}}_{-}
-2\off{\R_{F},\D\Psi_{T}}_{-}\nonumber\\
&=&2\off{\of{\D\R_{F}}\wedge\Psi_{T}-
\of{-1}^{\mf{deg}\Psi_{T}}\Psi_{T}\wedge\D\R_{F}},
\eeq
and the foliar counterpart of Bianchi's identity, 
\be{S172}
\D\R_{F}=0,
\ee
follows immediately. Of course, this result follows directly also from
eqs. (\ref{S96}) and (\ref{S162}).
%-=-=-=-=-=-=-=-=-=
\section{\sf The BRST super structure of the foliar bundle}\label{S3}
%-=-=-=-=-=-=-=-=-=
Our present aim is to explore the geometry induced along the vertical
directions (those that are parallel to the fibers). Consider the
following set of $m$ mutually-independent {\em horizontal\/} deformations 
\be{S180}
\ba{rcl}
\omega_{\alpha}\;\rightarrow\;\omega_{\alpha}+\Omega_{\alpha},
&&\alpha=1,\ldots,m,
\ea
\ee 
where the
deformation elements $\offf{\Omega_{\alpha}}$ are linear with respect
to all the $G$'s. Consequently, $\omega_{\alpha}+\Omega_{\alpha}$
transforms according to (\ref{S40}), but we require that
$\omega_{\alpha}$ and $\omega_{\alpha}+\Omega_{\alpha}$ cannot be
connected through a gauge transformation; in other words, 
the deformations display bijections between gauge-inequivalent
orbits. In general, the shifted connections correspond
to a different foliar curvature. But this may be avoided according
to the following prescription: One extends the basespace sector of
the bundle such that it includes also the angles associated with 
the gauge groups, and treats these angles as if they were additional
independent variables. One may then demand that the original curvature
remains inert, but then he must pay a price in the form of additional
structural constraints associated with the vertical directions.\Par  
Each set of angles $\offf{\phi^{a_{\alpha}}\of{x}}$, which coordinating
$G_{\alpha}\of{x}$, 
is naturally supplied with a coboundary-type operator $\delta_{\alpha}$,
in complete analogy with the exterior derivative $d$ on $M$. In other
words, each group $G_{\alpha}\of{x}$, at any $x\in M$, associates a
Grassmann space graded by $\delta_{\alpha}$. The differentiation with 
respect to an angle satisfies:
\beq{S182}
\frac{\delta\phi^{b_{\alpha}}}{\delta\phi^{a_{\alpha}}}=
\delta^{a_{\alpha}b_{\alpha}},
&\mbox{or more generally,}&
\frac{\delta\phi^{b_{\gamma}}}{\delta\phi^{a_{\alpha}}}=
\delta_{\alpha\gamma}\delta^{a_{\alpha}b_{\gamma}}
\eeq
because angles associated with different groups are independent of each
other whatsoever. The coboundary operator $\delta_{\alpha}$ is 
explicitly defined through
\beq{S184}
\delta_{\alpha}\phi^{a_{\alpha}}:=\delta\phi^{a_{\alpha}}
&\Rightarrow&
\delta_{\alpha}\equiv\delta\phi^{a_{\alpha}}
\frac{\delta}{\delta\phi^{a_{\alpha}}}.
\eeq
Over the extended space of differential forms, 
\be{S186}
\Upsilon=\Lambda^{n}\bigwedge_{\alpha=1}^{m}\Lambda^{n_{\alpha}},
\ee
$\delta\phi^{a_{\alpha}}\wedge{d}x^{\mu}=
-{d}x^{\mu}\wedge\delta\phi^{a_{\alpha}}$, %and
$\delta\phi^{a_{\alpha}}\wedge\delta\phi^{a_{\gamma}}=-
\delta\phi^{a_{\gamma}}\wedge\delta\phi^{a_{\alpha}}$,
$\alpha,\gamma=1,\ldots,m$,
$a_{\alpha\of{\gamma}}=1,\ldots,n_{\alpha\of{\gamma}}$, 
$\mu=1,\ldots,n$; hence all exterior derivatives anti-commute,
$d\delta_{\alpha}+\delta_{\alpha}d=\delta_{\alpha}\delta_{\gamma}
+\delta_{\gamma}\delta_{\alpha}=0$. Consequently, the $m$ pairs
$\of{d,\delta_{\alpha}}$ which act on the $m$ slices
$\Lambda^{n,n_{\alpha}}\subset\Upsilon$ 
give rise to $m$ bi-complexes of the form,
\be{S190}
\ba{cclclcc}
&&\vdots&&\vdots&&\\
&&\uparrow\delta_{\alpha}&&\uparrow\delta_{\alpha}&&\\
\cdots&\stackrel{\mbox{$d$}}{\rightarrow}&\Lambda^{i,j_{\alpha}+1}&
\stackrel{\mbox{$d$}}{\rightarrow}&
\Lambda^{i+1,j_{\alpha}+1}&\stackrel{\mbox{$d$}}{\rightarrow}&\cdots\\
&&\uparrow\delta_{\alpha}&&\uparrow\delta_{\alpha}&&\\
\cdots&\stackrel{\mbox{$d$}}{\rightarrow}&\Lambda^{i,j_{\alpha}}&
\stackrel{\mbox{$d$}}{\rightarrow}&
\Lambda^{i+1,j_{\alpha}}&\stackrel{\mbox{$d$}}{\rightarrow}&\cdots\\
&&\uparrow\delta_{\alpha}&&\uparrow\delta_{\alpha}&&\\
&&\vdots&&\vdots&&
\ea
\ee
where $0\leq{i}\leq{n}$, and $0\leq{j_{\alpha}}\leq{n_{\alpha}}$.\Par
Letting all of our bundle objects, in particular the connection
1-forms and the deformation terms, depend also on all group angles, 
requires a re-formulation of the bundle's
covariant exterior derivative:
$\D\rightarrow\widehat{\D}\,$, where
\beq{S200}
\widehat{\D}\Psi_{T}&:=&
d\Psi_{T}\;+\:\sum_{\alpha=1}^{m}\left(\,\delta_{\alpha}\Psi_{T}+
\omega_{\alpha}\wedge\Psi_{T}\;+\:
\of{-1}^{\ms{deg}\of{\Psi_{T}}+1}
\Psi_{T}\wedge\omega_{\alpha}\right.\nonumber\\
&&+\;\left.\Omega_{\alpha}\wedge\Psi_{T}
\;+\;\of{-1}^{\ms{deg}\of{\Psi_{T}}+1}
\Psi_{T}\wedge\Omega_{\alpha}\right).
%\\&=&\D\Psi_{T}+\sum_{\alpha}\of{\delta_{\alpha}\Psi_{T}+\Omega_{\alpha}\wedge\Psi_{T}+\of{-1}^{\mf{deg}\of{\Psi_{T}}+1}\Psi_{T}\wedge\Omega_{\alpha}}.
\eeq
In particular, and after doing some annoying algebra, two successive
applications of $\widehat{\D}$ on a generic $\Psi_{T}$ yields:
\beq{S210}
\widehat{\D}\widehat{\D}\Psi_{T} &=& \off{\R,\Psi}_{-}+\off{\sum_{\alpha}
\D\Omega_{\alpha},\Psi_{T}}_{-}+\off{\sum_{\alpha,\gamma}
\delta_{\gamma}\omega_{\alpha},\Psi_{T}}_{-}\nonumber\\
&& +\,\off{\sum_{\alpha,\gamma}
\delta_{\gamma}\Omega_{\alpha},\Psi_{T}}_{-}+
\off{\sum_{\alpha,\gamma}\Omega_{\alpha}\wedge\Omega_{\gamma},\Psi_{T}}_{-}.
\eeq
($\D$ in $\D\Omega_{\alpha}$ is the covariant exterior derivative with 
respect to the original reduced base).\Par
We shall now associate the deformation elements $\offf{\Omega_{\alpha}}$
with ghost fields (the $\delta$'s turn out to generate ghost numbers -
see eq. (\ref{S300})). This is suggested by the following argument: 
If we now require that the curvature $\R_{F}$ remains inert during the 
extension of the bundle, then the extra four terms in eq. (\ref{S210})
must sum up to zero. Comparing terms of equal ``{\em ghost
grading\/}'' we find the following variation laws to hold:   
\beq{S220}
\delta_{\left[\alpha\right.}\Omega_{\left.\gamma\right]_{+}}&\!=\!&
-\Omega_{\left[\alpha\right.}\!\wedge\Omega_{\left.\gamma\right]_{+}}\\
\delta_{\alpha}\of{\sum_{\gamma=1}^{m}\omega_{\gamma}}
&=&-\D\,\Omega_{\alpha}\,;
\label{S230}
\eeq
without loss of generality we pick for (\ref{S220}) the variation law
$\delta_{\alpha}\Omega_{\gamma}=-\Omega_{\gamma}\wedge\Omega_{\alpha}$.
One now easily verifies that $\delta_{\alpha}$
squares to zero on $\Omega_{\gamma}$: 
\be{S240}
\delta_{\alpha}\delta_{\alpha}\Omega_{\gamma}=
-\of{\delta_{\alpha}\Omega_{\gamma}}\wedge\Omega_{\alpha}+
\Omega_{\gamma}\wedge\delta_{\alpha}\Omega_{\alpha}=
\Omega_{\gamma}\wedge\Omega_{\alpha}\wedge\Omega_{\alpha}-
\Omega_{\gamma}\wedge\Omega_{\alpha}\wedge\Omega_{\alpha}=0,
\ee
and also on $\omega_{F}=\sum_{\gamma}\omega_{\gamma}$ (recall formula
(\ref{S160})): 
\beq{S250}
\delta_{\alpha}\delta_{\alpha}\omega_{F}&=&
-\delta_{\alpha}\of{d\Omega_{\alpha}
+\omega_{F}\wedge
\Omega_{\alpha}+\Omega_{\alpha}\wedge\omega_{F}}
\nonumber\\&=&
%-d\of{\Omega_{\alpha}\wedge\Omega_{\alpha}}
%+\D\Omega_{\alpha}\wedge\Omega_{\alpha}
%-\omega_{F}\wedge
%\Omega_{\alpha}\wedge\Omega_{\alpha}+
%\Omega_{\alpha}\wedge\Omega_{\alpha}\wedge
%\omega_{F}-
%\Omega_{\alpha}\wedge\D\Omega_{\alpha}\nonumber\\&=&
-\D\of{\Omega_{\alpha}\wedge\Omega_{\alpha}}+
\D\Omega_{\alpha}\wedge\Omega_{\alpha}-
\Omega_{\alpha}\wedge\D\Omega_{\alpha}\;=\;0.
\eeq
At this point we already see why it is suggestive to associated the
shifts at the gauge sector with ghosts: Eqs. (\ref{S220})-(\ref{S230})
constitute the BRST algebra associated with the folium $F$. Note that
the sum as a whole, 
$\omega_{F}=\sum_{\gamma}\omega_{\gamma}$, and not each particular
summand, possesses a definite transformation law. Hence, each
$\delta$-variation detects a {\em single-gauge\/} connection despite
the fact that many symmetry structures are involved in our bundle
construction.\Par 
The coboundary operators $\offf{\delta_{\gamma}}$ can also be
interpreted as those operators that generates the deformation elements
when applied to the multi-foil basis:
\be{S300}
\delta_{\gamma}\bs{e}_{A_{1}\cdots A_{m}}=
-\rho_{F}\of{\Omega_{\gamma}}^{B_{1}\cdots B{m}}_{A_{1}\cdots A_{m}}
\bs{e}_{B_{1}\cdots B_{m}}.
\ee
On the basis of this definition we may directly re-derive the
constraint (\ref{S220}):
\be{S310}
\ba{rcl}
0=\of{\delta_{\alpha}\delta_{\gamma}
+\delta_{\gamma}\delta_{\alpha}}\bs{e}=
-\of{\delta_{\left[\alpha\right.}\Omega_{\left.\gamma\right]_{+}}
+\Omega_{\left[\gamma\right.}\!\wedge\Omega_{\left.\alpha\right]_{+}}}\bs{e}
&\Rightarrow&
\delta_{\left[\alpha\right.}\Omega_{\left.\gamma\right]_{+}}=
-\Omega_{\left[\gamma\right.}\!\wedge\Omega_{\left.\alpha\right]_{+}}\,,
\ea
\ee
and the constraint (\ref{S230}):  
\beq{S320}
0&\!=\!&\of{\delta_{\gamma}d+d\delta_{\gamma}}\bs{e}=
-\delta_{\gamma}\of{\omega_{F}\bs{e}}
-d\of{\Omega_{\gamma}\bs{e}}\nonumber\\
%&\!=\!&
%-\of{\delta_{\gamma}\omega_{F}}\bs{e}-
%\omega_{F}\wedge\Omega_{\gamma}\bs{e}
%-\of{d\Omega_{\gamma}}\bs{e}-\Omega_{\gamma}\wedge
%\omega_{F}\bs{e}
%\nonumber\\
&\!=\!&\of{-\delta_{\gamma}\omega_{F}-\D\Omega_{\gamma}}\bs{e}
\;\;\Rightarrow\;\;\delta_{\gamma}\omega_{F}=-\D\Omega_{\gamma}.
\eeq
In particular, eq. (\ref{S310}) ($\equiv$(\ref{S220})) is a
generalization of the Maurer-Cartan equation(s) to foliar bundles;  
the `off-diagonal' equations correspond to cross-fiber
interferences.\footnote{Over the bundle whose basespace is enlarged,
the two definitions, (\ref{S100}) and (\ref{S300}), can be combined
into 
\[
\of{d+\sum_{\alpha=1}^{m}\delta_{\alpha}}\bs{e}_{A_{1}\cdots{A}_{m}}
=-\sum_{\alpha=1}^{m}\rho_{F}\of{\widetilde\omega_{\alpha}+
\widetilde\Omega_{\alpha}}\!_{A_{1}\cdots{A}_{m}}^{B_{1}\cdots{B}_{m}}
\bs{e}_{B_{1}\cdots{B}_{m}},
\]
which we may abbreviate as
$\of{d+\delta_{F}}\bs{e}=-\of{\omega_{F}+\Omega_{F}}\bs{e}$. Then, the 
modified covariant exterior derivative of $\Psi_{T}$ (formula
(\ref{S200})) takes the succinct form,
\[
\widehat\D\Psi_{T}=
\of{d+\delta_{F}}\Psi_{T}+\gb{\omega_{F}+\Omega_{F}}{\Psi_{T}}.
\]}\Par
A prior knowledge of the extended Maurer-Cartan equations pins down 
an equivalent (but not self-contained) description for the ghost sector:
Let us define the 2-form quantities, 
$B_{\alpha\gamma}=\delta_{\alpha}\Omega_{\gamma}
=-\Omega_{\gamma}\wedge\Omega_{\alpha}$, $\alpha\neq\gamma$. 
Now, the nilpotency of $\delta_{\alpha}$ reads:
$\delta_{\alpha}B_{\alpha\gamma}=0$; on the other hand,
$\delta_{\gamma}B_{\alpha\gamma}=-\gb{\Omega_{\gamma}}{B_{\alpha\gamma}}
=-\off{\Omega_{\gamma},B_{\alpha\gamma}}_{-}$, hence
$\delta_{\gamma}\delta_{\gamma}B_{\alpha\gamma}=0$. According to these
variation laws, we are dealing here with the $B$-fields associated
with the BRST symmetry on $F$. However, in contrast with the 
traditional treating \cite{B,NO}, these $B$-fields are by no means
auxiliary degrees of freedom; rather, they 
are composites made of ghost-ghost pairs.\Par 
The variation laws (\ref{S220})-(\ref{S230}) are manifestly invariant 
under a duality transformation which is realized by interchanging of
labels, $\alpha\leftrightarrow\gamma$, applied
simultaneously to both equations. As for the $B$-fields, the duality
manifests itself via transposition. This provides us with an {\em
arbitrary\/} classification into ghost-antighost pairs, and with
the corresponding pairs of BRST and anti-BRST operators.\Par 
Consider for example the 2-folia case $\of{\alpha,\gamma=1,2}$, and put 
$\delta_{1}=\delta$, $\delta_{2}=\bar\delta$, $\Omega_{1}=\Omega$,
$\Omega_{2}=\Phi$, $\omega_{1}=\omega$, and $\omega_{2}=\varphi$.
Then, from formulas (\ref{S220}) and (\ref{S230}), we have
\be{S340}\ba{ccccc}
\delta\Omega=-\Omega\wedge\Omega&&\delta\Phi=-\Phi\wedge\Omega &&
\delta\of{\omega +\varphi}=-\D\Omega\\
\bar\delta\Phi=-\Phi\wedge\Phi&&\bar\delta\Omega=-\Omega\wedge\Phi &&
\bar\delta\of{\omega +\varphi}=-\D\Phi. \label{Sb34}
\ea\ee
In particular, $\delta\Phi+\bar\delta\Omega=-\gb{\Phi}{\Omega}$.
A simultaneous exchange, 
\beq{S344}
\delta\;\leftrightarrow\;\bar\delta,&&
\Omega\;\leftrightarrow\;\Phi,
\eeq
transforms the upper triad in (\ref{S340}) into 
the lower one, and vice versa.
Otherwise, we may set $B=-\Phi\wedge\Omega$ and $\bar B=-\Omega\wedge\Phi$,
whose variation properties are easily read-off from (\ref{S340}),
\be{S350}\ba{lcl}
\delta{B=0}&&\delta\bar{B}=-\bgb{\Omega}{\bar{B}}\\
\bar\delta\bar{B}=0&&\bar\delta{B}=-\gb{\Phi}{B}
\ea\ee
(whence $\delta\bar\delta B$ and $\bar\delta\delta\bar B$ vanish
independently). The duality transformation (\ref{S344})
maps a $B$-field into its dual $\bar B$, and the upper pair in
(\ref{S350}) is mapped into the lower one. 
\Par\Par\Par\noindent
%=-=-=-=-=-=-=---=--=-=-=-=-=-=
{\Large{\sf{Acknowledgments}}}:
%=-=-=-=-=-=-=---=--=-=-=-=-=-=
I thank Oded Kenneth for a helpful discussion, Yuval Ne'eman and Larry
Horwitz for commenting on the manuscript, and Jim Stashef for making 
constructive comments on the early version of this
work. This work was partially supported by the Ann \& Maurice 
Jacob Cohen Doctoral Fellowship in Nuclear and Particle Physics.

\end{document}